\documentclass[aps,prb,twocolumn,floatfix]{revtex4}
\def\beq{\begin{equation}}
\def\eeq{\end{equation}}
\def\beqn{\begin{eqnarray}}
\def\eeqn{\end{eqnarray}}
\newcommand{\bkbox}{(Ba$_{1-x}$K$_{x}$)BiO$_{3}$}

\usepackage{graphicx}
\begin{document}

%\newlength\FigWidth
%\FigWidth 3.25 true in
%\FigWidth 5 true in
\draft
\title{Spatial Structures in a Generalized Ginzburg-Landau Free Energy}
\author{Abdel-Khalek Farid$^\ast$, Yonggang Yu$^\ast$, Avadh Saxena$^\dag$ and Pradeep Kumar$^\ast$} 
\affiliation{$^\ast$Department of Physics, PO Box 118440, University of Florida, 
Gainesville, Florida 32611-8440}
\affiliation{$^\dag$Theoretical Division, Los Alamos National Laboratory, 
Los Alamos, New Mexico 87545}
\date{\today}
\begin{abstract}

Searching for characteristic signatures of a higher order phase transition (specifically of order three or four), we have calculated the spatial profiles and the energies of a spatially varying order parameter in one dimension.  In the case of a $p^{th}$ order phase transition to a superconducting ground state, the free energy density depends on temperature as $a^p$, where $a = a_o(1-T/T_c)$ is the reduced temperature.  The energy of a domain wall between two degenerate ground states is $\epsilon_p \simeq a^{p-1/2}$.  We have also investigated the effects of a supercurrent in a narrow wire.  These effects are limited by a critical current which has a temperature dependence $J_c(T) \simeq a^{(2p-1)/2}$.  The phase slip center profiles and their energies are also calculated.  Given the suggestion that the superconducting transtion in \bkbox, for $x = 0.4$, may be of order four, these predictions have relevance for future experiments.  
\end{abstract}
%\pacs{}
\maketitle

\section{Introduction}

A large part of statistical mechanics, having to do with the physics of phase transitions and nonlinear properties is based on a rather fundamental generalization, due to Landau\cite{landausm,mattis}, of the well-known Gibbs free energy.   There is a functional $\tilde{F}$ depending on an (in general complex) ``order parameter" $M({\bf r})$, given by,  

\beq
\label{eq12}
\tilde{F}[{\bf M}({\bf r})] = \int d{\bf v} [ - a |{\bf M}|^2 + b |{\bf M}|^4 + c |\nabla {\bf M}|^2 - {\bf M} \cdot {\bf H}], 
\eeq

where $a \propto (1-T/T_c)$ and $b$ and $c$ are nonnegative, temperature independent constants.  Here $\bf H$ represents an external field which couples directly to the order parameter.

This free energy is the backbone of a vast literature\cite{kadanoff} on second order phase transitions, critical phenomena as well as nonlinear properties.  Regardless of the microscopic fabric of the system under consideration, near a second order phase transition where universal features appear which are independent of the details of interaction, this free energy contains all of the necessary information.  Thus a superconductor or a magnet or a nematic liquid crystal, all have the same Ginzburg-Landau (GL) free energy as long as the appropriate order parameter $M$ is identified.  It is however limited to describe only the vicinity of a second order phase transition.  When the fluctuations are ignored, the corresponding Gibbs free energy $G$, identified as the minimum of the functional $\tilde{F}$, has the temperature dependence of $G \simeq a^2$.  

According to the classification of phase transitions as proposed by Ehrenfest\cite{ehren}, the transitions in general can be of any order.  In a second order phase transition, the specific heat and the compressibility, which are the second order derivatives of the free energy (with respect to temperature and pressure respectively), are discontinuous at the transition.  In general though, the discontinuity is often replaced by a weak, often logarithmic singularity.  One could then view the Ehrenfest definition of an order as one where the lower derivatives of the free energy are continuous at the transition but the higher derivatives are singular.   Thus in a fourth order phase transition, all third order derivatives are continuous, all fifth order derivatives are singular at the transition.  The fourth order derivatives are either discontinuous or are weakly singular.  

The generalization of the Landau free energy to a higher order phase transition is achieved via a weight function.  For a third and a fourth order phase transitions, we need\cite{k1,k2,ks} respectively, within mean field in the sense discussed above, $G \simeq a^3$ and $G \simeq a^4$.  This can be obtained from ``Landau-like" free energies defined with a weight function:
\beq
\label{eq13}
 F_{III} = \int dv |M|^2[ - a_3 |M|^2 + b_3 |M|^4 + c_3 |\nabla M|^2], 
\eeq
\beq
\label{eq14}
  F_{IV} = \int dv |M|^4[ - a_4 |M|^2 + b_4 |M|^4 + c_4 |\nabla M|^2]. 
\eeq
In general the weight function is given by $|M|^{2(p-2)}$ for $p\ge2$. Here $a_p$'s change sign at $T_c$  and are often used below for the reduced temperature $a_p = {a_p}^o(1-T/T_c)$.  The other constants $b_p$ and $c_p$, $p = 3,4$ are non-negative and generally temperature independent.  In a charged fluid (such as a superconductor\cite{schrieffer, tinkham}), the coupling to a magnetic field takes the form of a gauge transformation $\nabla \rightarrow \nabla-2\pi iA/\phi_o$.  Here $\phi_o = h/2e$ is the superconducting flux quantum.  The corresponding Gibbs free energy is to be identified with the minimum of the functional $F$ above with respect to the order parameter.  It has the expected temperature and field dependence.

The importance of these free energies is highlighted by the recent discovery\cite{hall,k1} of anomalies in the superconducting transition in \bkbox  (x = 0.4).  These anomalies, a missing discontinuity in specific heat\cite{hundley, note1} as well as in susceptibility\cite{hall,k1}, indicate a phase transition of order higher than second.  The actual order in BKBO has been determined\cite{k1} to be fourth.  In another example, the specific heat\cite{junod} in Bi2212 (chemical compound)has been noted to be kink-like, representing a third order phase transition.  The tell-tale signature of a third order phase transition, namely a penetration depth $\lambda^{-2} \simeq a^2$ remains to be investigated.  There are other examples in Ref. [\onlinecite{k2}] where some of the anomalies have been seen but a detailed study still needs to be carried out.  Regardless of the detailed microscopic reasons, the free energies in Eqs. (\ref{eq13},\ref{eq14}) characterize the phase transition in its entirety.  For BKBO, the function M is the superconducting order parameter.  In the absence of a microscopic theory we do not know whether it is also the energy gap at the Fermi surface, although the tunnelling measurements\cite{tunnelling} seem to be consistent with this identification.  The earlier analysis\cite{k1,k2,ks} has been based on a uniform order parameter, except for some straightforward effects of a magnetic field such as a derivation of the London equation and the temperature dependence of the penetration length.

This paper is aimed at studying the properties of a substantially inhomogeneous order parameter such as found\cite{nelson,mineev,vw}  in defects and textures in the condensed state.  Their analysis is sometimes carried out within homotopy theory\cite{mineev} where the geometrical/topological aspects can be studied in detail.  These (geometrical/topological) properties, characteristic of a mapping between the real space and the order parameter space, will be unaffected by the order of the transition.  All quantitative properties \cite{nelson} for example the energies of defects, how do two defects interact or the effects of the defects and textures on the thermodynamic properties; they will all be different.  In this paper we are interested in the solutions of the relevant nonlinear differential equation.  To begin with, let us limit ourselves to one space dimension. The defects discussed here are domain walls in three dimensions (3D) or topological point defects that occur in narrow wires.  In physical terms, these are the energies and the profiles of the order parameter between two degenerate superconducting states.  A special case in Sec. IV below refers to phase slip centers.  Voltage drops appear across these objects in a narrow wire in the presence of a current. 

To recall the results of an earlier analysis\cite{k1,k2,ks} for a $p^{th}$ order phase transition:  (1) The thermodynamic free energy at the order parameter minimum follows $F_{op} \propto a^p$.  (2) The superfluid density satisfies $ \rho_s \propto a^{p-1}$.  This leads to the result that $\lambda^{-2} \simeq H_{c1} \propto a^{p-1}$, where $\lambda$ is the London penetration depth.  (3) Since the coherence length $\xi$, which measures the stiffness of the order parameter, has the temperature dependence $\xi^{-2} \propto a$, the Landau parameter $\kappa = \lambda/\xi$ is temperature dependent (in contrast to a superconductor undergoing a second order phase transition where $\kappa$ is a temperature independent constant) and follows, $\kappa \propto a^{1-p/2}$. 
 
The principal results in this paper are:  When a defect is created, the order parameter is suppressed in a small region of order of the coherence length, $\xi$.  If the bulk condensation energy density (energy per unit length) is $E_o$, the energy of a defect is $\propto E_o \xi $.  This has the temperature dependence of $a^{p-1/2}$.  The numerical factor in front of this expression is calculated below in detail for some specific cases.  The critical current in a narrow wire is known to be $J_c (T) \propto a^q$ with $q = 3/2$ for a second order phase transition.  Below we derive this exponent for a third and a fourth order phase transition, where respectively $q = 5/2$ and $7/2$.  For a $p$th order phase transition, the exponent can be estimated as follows.  Consider the free energy of  a narrow wire in the presence of a current $J$; it must be $ F \simeq a^p \simeq J^2/{2 \rho_s}$.  It follows readily that given the temperature dependences described above, $q= (2p-1)/2$. The energy of a phase slip center, and therefore the temperature dependence of resistivity near a suprconducting phase transition was first calculated by Langer and Ambegaokar\cite{la, tinkham}.   A thermal distribution of these localized  voltage points leads to an activated/exponential rise in resistance with the activation energy being the energy of a single phase slip center.  The latter in turn is the above mentioned energy of a defect, of the order of  $\propto E_o \xi $.

The outline of this paper is as follows.  Section II contains the mathematical formalism we have used.  In particular this section contains the Euler-Lagrange equation for a one dimensional nonlinear field theory derived from Eqs. (\ref{eq13},\ref{eq14}).  In Sec. III, we describe the solutions of the nonlinear partial differential  equations.  The current induced effects are discussed in Sec. IV.  These include temperature dependence of the critical current in a narrow wire and the temperature dependence of the resistance near the phase transition.   Finally the last section contains a summary of our conclusions.

\section{Mathematical Infrastructure}

The gradient terms in Eqs. (\ref{eq13},\ref{eq14}) contain information about the stiffness of the order parameter.  The precise degree of this stiffness, represented by the coefficients and which can be measured in a superconductor in either the upper critical field (the resistance of the order parameter to spatial variation) or in the London penetration depth (and the lower critical field, measuring the resistance to magnetic field), depends on the material parameters.   The limitation to the lowest order gradient terms is initially motivated by an esthetic curiosity about the long wavelength phenomena.  Eventually though, the ability of this formalism to describe defects and textures is an a posteriori rationalization.

An Euler-Lagrange equation for the free energy (for a scalar order parameter $M$) described in Eq. (\ref{eq13}) is given by 
\beq
\label{eq21}
0  = - 2 a_3 M^3 +3 b_3 M^5 - c_3[ M(\nabla M)^2 + M^2(\nabla^2M)]. 
\eeq

In one dimension (1D), $\nabla = \partial_z$ and $\nabla^2 = \partial_{zz}$.  The letter subscripts denote differentiation.  The dimensional variables can be scaled out with the transformations
\beq
\label{eq22}
M = \sqrt{{2a_3}\over{3b_3}}n;~~~~ y = z/\xi_3;~~~~ {\xi_3}^2 = c_3/2a_3; 
\eeq
Eq. (\ref{eq21}) then becomes
\beq
\label{eq23}
n(n_y)^2 + n^2n_{yy} +n^3 - n^5 = 0.
\eeq
The function $f = n^2$ satisfies

\beq
\label{eq24}
-{1 \over 2}f_{yy} -f +f^2 =0. 
\eeq

The corresponding energies are then calculated to be,
\beq
\label{eq221}
F_{III} = E_3 \epsilon_3 ;~~~~ E_3 = \xi_3 {{(2a_3)^3}\over{(3b_3)^2}}, 
\eeq
and 
\beq
\label{eq241}
\epsilon_3 ={1 \over 6} \int_0^1 dy \left[ {3 \over 2}\left({\partial f \over \partial y}\right)^2+1 - 3f^2 + 2f^3\right]. 
\eeq
Since the defects are localized over a spatial region of order $\xi_3$, we notice that the energy scale follows the temperature dependence $a_3^3\xi_3$.  The function n(y) approaches $\pm 1$ in the bulk.  The dimensionless energy expression has been adjusted to define the defect energy with respect to the state with a uniform order parameter.

The functional $F_{III}[f(y)]$ has to be limited to $f(y) \ge 0$.  If it were not so, then Eq.(\ref{eq241}) as a functional of $f$ would be unbounded from below and the mean field theory (where the sum over possible configurations is believed dominated by the minimum energy configuration) is then undefined.  Since $f = n^2$, the condition is also a natural requirement.
 
The physics derived from Eq. (\ref{eq14}) for a fourth order transition is similar, although analytically less tractable.  The Euler-Lagrange equation, in this case, becomes,

\beq
\label{eq25}
c_4[M^4M_{zz}+2M^3(M_z)^2] + 3a_4M^5 -4b_4M^7 = 0. 
\eeq

Following very similar transformations (removal of dimensional variables), but with an important difference, namely that $f \propto M^3$, the equation corresponding to Eq. (\ref{eq24}) becomes;

\beq
\label{eq26}
{1 \over 3}f_{yy} +f - f^{5/3} = 0 ,  
\eeq
where $f^{1/3}= n = M/\sqrt{3a_4 \over 4b_4}$ and $y = z/\xi_4$ with ${\xi_4}^2 = c_4/{3a_4}$.  Also, the energy of these textures is given by $F_{IV} = E_4\epsilon_4$ with $E_4 = \xi_4 (3a_4)^4/(4b_4)^3$.  The quantity $\epsilon_4$, as an integral, is given by 
\beq
\label{eq27}
\epsilon_4 = {1 \over 12} \int dy \left[1- 4f^2 + 3f^{8/3} + {4 \over 3}\left({\partial f \over \partial y}\right)^2\right]. 
\eeq

\section{Spatial Configurations}
In this section we calculate the spatial profiles.  This is done separately for the third and fourth order free energies.  The solutions are for a scalar order parameter in one dimension.

\subsection{Third Order Free Energy}
In case of Eqs. (\ref{eq13} and \ref{eq23}) the transformation $f = n^2$ eliminates a sign degeneracy.  It also restricts $f \geq 0$.  A domain wall between regions with $n = 1$ and $n = -1$, the two degenerate ground states in bulk, becomes a profile where $f = 1$ everywhere, except in a narrow region of order $\xi$, where it vanishes.   Let us first consider the solution for Eq. (\ref{eq24}).  A first integral is obtained by multiplying  the partial differential equation by $n_y$ and integrating over y.  This leads to the first integral,
\beq
\label{eq31}
{1 \over 4}(f_y)^2+{f^2 \over 2} - {f^3 \over 3} = K. 
\eeq

Recall that the traditional analysis from here onwards, notices,  that if $y$ were time, $K$ would be the energy of a particle with position $f$, moving in a potential well $-V$, where $V$ is the gradient-free part of the free energy (for $f \geq 0$).  For small $K$, the solutions are periodic, similar to a periodic arrangement of holes in the condensate.  Equation (\ref{eq31}) may be integrated in terms of elliptic functions. As $K$ increases, the sinusoidal functions sharpen into square wave like structures, their wavelength increases until eventually for $K \rightarrow K_0 = 1/6$, one obtains a solitary wave solution.  (In general, $K_0 =1/2p$.)  The center of the solitary wave in $f(y)$ is linear.  Since at the center the order parameter is small it follows that $f(y) \simeq 2\sqrt{K} y$.

For small $K$, the potential for the motion of the fictitious particle is simply $f^2$.  The motion in a quadratic well is described by $f= \sqrt{2K}|\sin (y\sqrt{2})|$.   Since $f \geq 0$, the analytic solution needs to be understood in light of this positivity constraint. 

The energy of the small amplitude periodic structure (up to order linear in $K$) is zero, as can be evaluated by direct calculation.  As $K$ increases and the periodic structure begins to resemble a soliton anti-soliton lattice with wide separation between the kinks, the energy reduces to the sum of the rest masses of the individual kinks, reduced by the attractive interaction between them.  For large distances $d$ between two domain walls, the asymptotic interaction between a soliton and an antisoliton\cite{manton} is given by 
\beq 
\label{eq326}
U(d)=-36\sqrt{2}~e^{-\sqrt{2}d} . 
\eeq 
                   
\begin{figure}[t]
\centering
\includegraphics[width=80mm]{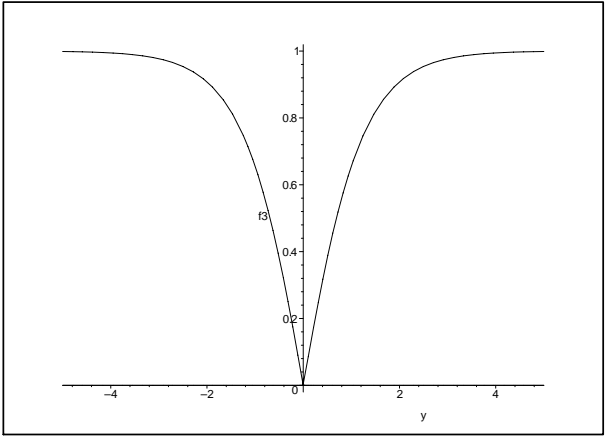}
%\hspace{0.0cm}\psfig{figure=f3.eps,height=6.5cm,width=6.5cm,angle=-90}
\caption{The function $f(y)$ as a function of $y$ for a third order free energy.}
\label{fig1}
\end{figure}

Integrating Eq.(\ref{eq31}) for $K=1/6$ we get  
 
\beq
\label{eq32}
f = {1 \over 2}\left[3 \tanh^2\left(\frac{y+X}{\sqrt{2}}\right) -1\right]. 
\eeq
Here X is the second integration constant, it represents the translational freedom in locating the center of the solitary wave anywhere.  Since $f \geq 0$, it follows that $X \geq X_c = \sqrt{2}\tanh^{-1}{1 \over \sqrt{3}} = 0.93$.  The solution for $y <0$ is obtained by folding the result around the vertical axis. Figure.(\ref{fig1}) shows the solution for $f(y)$.  This is different from a solution one obtains for a conventional solitary wave as in Eq. (\ref{eq12}), for a second order transition.  There is a discontinuity of slope at $y=0$ which is remnant of the similar discontinuity in the small amplitude solutions.  It arises from the constraint $f > 0$.  
  
\begin{figure}[b]
\centering
\includegraphics[width=80mm]{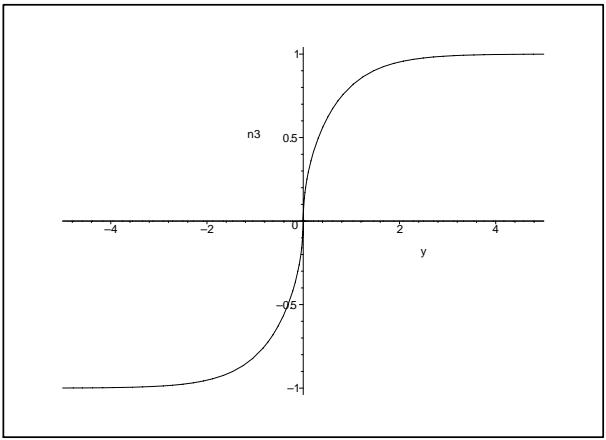}
%\hspace{0.0cm}\psfig{figure=n3.eps,height=6.5cm,width=6.5cm,angle=-90}
\caption{Domain wall profile $n(y)$ for a third order phase transition.
}
\label{fig2}
\end{figure}

The solution of the original Eq.(\ref{eq23}) is obtained from $n = \sqrt{f}$ with the positive root $y>0$ and the negative root for $y<0$.  The function $n(y$ is shown in Fig.(\ref{fig2}).

The energy of the solitary wave structure is given by 
\beq
\label{eq33}
\epsilon_3 = {3\sqrt2 \over 5}\left(1-{2 \over {3\sqrt3}}\right) = 0.522. 
\eeq

The $n(y)$ solution is depicted in Fig. (\ref{fig2}).  This is generated from the solution for $f(y)$ shown in Fig. (\ref{fig1}) by taking the positive square root for $y \geq 0$ and the negative square root for $y \leq 0$.  Near the center the function $n(y) \propto \sqrt{y}$.   

\subsection{Fourth Order Free Energy}

\begin{figure}[t]
\centering
\includegraphics[width=80mm]{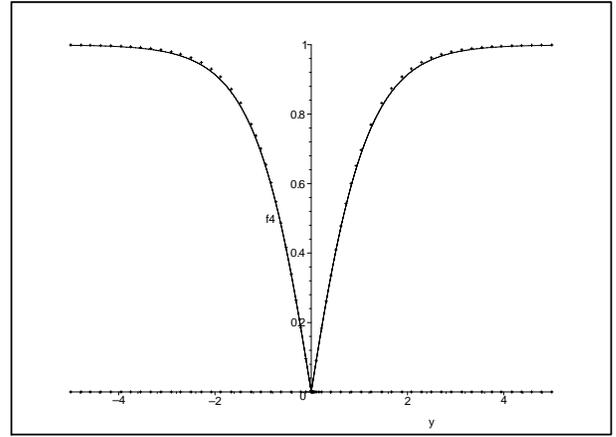}
%\hspace{0.0cm}\psfig{figure=f41.eps,height=6.5cm,width=6.5cm,angle=-90}
\caption{The approximate solution, Eq. (\ref{eq39}) (solid line), and numerically exact solution (crosses) of the function $f(y)=n^3$ for a fourth order phase transition, obtained from Eq.(\ref{eq35}}
\label{fig3}
\end{figure}

For a fourth order transition free energy, Eq. (\ref{eq14}), the Euler-Lagrange equation, Eq. (\ref{eq26}), can be readily integrated once.  The resulting expression is

\beq
\label{eq34}
{1 \over 6}({f_y})^2 +{1 \over 2}f^2 - {3 \over 8}f^{8/3} = K.
\eeq
Again for small $K$, one has the periodic solutions which start out as sinusoidal solutions but sharpen into square wave like functions as $K$ increases.  For $K = 1/8$, the solution for a single defect is given implicitly by the integral

\beq
\label{eq35}
\int_{f(0)}^f {d\overline{f} \over {\sqrt{1-4{\overline{f}}^2 
+ 3{\overline{f}}^{8/3}}}} = \frac{\sqrt{3}}{2}y. 
\eeq
The solution $f(y)$ with $f(0) = 0$, is sketched in Fig.(\ref{fig3}) as the set of crosses.  A numerical integration of Eq. (\ref{eq27}) leads to the value of the integral as $\epsilon_4 = 0.241$.  The defect energies get smaller for the higher order free energies.  The ratio $\epsilon_3/\epsilon_4 = 2.17$.  In comparison, the domain wall energy for Eq.(\ref{eq12}), $\epsilon_2 = 2\sqrt2/3 = 0.943$ and $\epsilon_2/\epsilon_3 = 1.81$. 

An approximate solution for Eqs. (\ref{eq25}, \ref{eq26}) can be worked out by starting with the free energy in dimensionless variables, as described in Eq.(\ref{eq27}).  This free energy can be broken into two parts; $\epsilon_4 = A + B$ where, 
\beq
\label{eq38}
A = {1 \over 12} \int dy \left[1- 3f^2 + 2f^3 + {4 \over 3}\left({\partial f \over \partial y}\right)^2\right], 
\eeq
and
\beq
\label{eq39}
B = {1 \over 12} \int dy [3f^{8/3} - f^2(1+2f)]. 
\eeq
The separation here is guided by the solvability.  Its validity is justified a posteriori.  The unperturbed part of the free energy, represented by $A$ is similar to Eq. (\ref{eq241}).  There is an overall factor of 2 ($\epsilon_4$ is smaller) and the size of the structure is slightly smaller by a factor $\sqrt{8/9}$.  The perturbation $B$ is non-zero only when $f \neq 1$.

The solution to Eq. (\ref{eq38}) is readily obtained.

\beq
\label{eq391}
f^o (y) = {1 \over 2}[3 \tanh^2(3(y+X)/4) -1]. 
\eeq
The energy of this structure is given by $ A = 0.246$, rather close to the exact value for the energy (0.241) cited following Eq. (\ref{eq35}).  The integral for $B$ can be evaluated using $f^o (y)$.  It is equal to $-0.001$.  Shown in Fig.(\ref{fig3}) as the solid line, is the solution described in Eq.(\ref{eq391}).  The difference between the exact numerically integrated solution and the approximate one is negligible.

\section{Current Flow Effects}

Consider a narrow wire (lateral dimensions smaller than the coherence length) made of a material described by the generalized free energies.  If the wire carries current $J$, the superfluid density is reduced, first quadratically but as the current reaches the critical current, more precipitiously.  The superfluid density vanishes at a critical current $J_c(T)$.  However even for $J \leq J_c(T)$, voltage drops appear in small regions called phase slip centers.  A comprehensive and lucid discussion of both the mathematical as well as physical effects has been provided by Langer and Ambegaokar\cite{la}.  

The physical effects remain qualitatively the same for a higher order phase transition.  The mathematical details, however, are different for the free energies in Eqs. (\ref{eq13}, \ref{eq14}).  In the presence of a constant current $J$, the free energy corresponding to Eq. (\ref{eq13}) for a third order transition becomes,

\begin{widetext}
\beq
\label{eq41}
\tilde{F_j} = F - J.A = \int dv |M|^2\left[ - a_3 |M|^2 + b_3 |M|^4 + c_3 |\nabla M|^2 - {{J^2} \over {4c_3M^6}}\right]. 
\eeq
\end{widetext}

Here $A$ is a vector potential, conjugate to the supercurrent $J$.  Because of the use here of the Legendre's transformation to develop $\tilde{F_j}$ as a function of specified $J$, the final equation is entirely in terms of the amplitude of the order parameter.  Its minimum yields the equilibrium order parameter, including the effects of a finite supercurrent $J$.

The physics described in Eq. (\ref{eq41}) begins with the usual free energy in Eq. (\ref{eq13}) for $J = 0$.  With increasing $J$, the order parameter $M$ corresponding to the minimum of the free energy moves to smaller values.  The free energy also has a maximum which separates the minimum from the normal state for $M=0$.  Both the minimum as well as the maximum disappear for $J \geq J_c(T)$ and there is no order parameter solution.  The minimum free energy order parameter solution $M(J)$ is a solution of the equation,

\beq
\label{eq42}
J^2 = c_3 M^5[4a_3 M^3-6b_3 M^5] . 
\eeq

For each $J < J_c$, there are two solutions.  The smaller M corresponds to the local maximum of the free energy.  The larger solution closer to the zero current equilibrium value is the global minimum and ceases to exist for $J > J_c(T)$.  The corresponding problem for Eq. (\ref{eq12}) leads to a temperature dependence of $J_c(T) \simeq a^{3/2}$.  For Eq. (\ref{eq13}), the result is $J_c(T) \simeq {a_3}^{5/2}$.  For Eq. (\ref{eq14}), the critical current has an even weaker temperature dependence, $J_c(T) \simeq {a_4}^{7/2}$.  It is straightforward to extend this calculation to an arbitrary order phase transition.  For a $p$th order transition the exponent for the temperature dependence is $(2p-1)/2$.  We thus have another characteristic signature of the order of a transition in the temperature dependence of the critical current in a wire. 

The solution corresponding to a phase slip center for Eq. (\ref{eq12}) was originally obtained\cite{la} by Langer and Ambegaokar.  In dimensionless variables, Eq. (\ref{eq41}) transforms into,

\beq
\label{eq43}
\tilde{F}_{III} =  E_3\int dy\left[ - {1 \over 2}f^2  + {1 \over 3}f^3 +  {1 \over 4}{f_y}^2 - {{g^2} \over {2f^2}}\right]. 
\eeq
The notation is similar to that following Eqs. (\ref{eq22}) and (\ref{eq221}) with the addition for the current.  Thus $g^2 = (J^2/2c_3)((3b)^4 / (2a_3)^5)$.  In terms of the dimensionless variables, Eq. (\ref{eq42}) becomes

\beq
\label{eq44}
g^2 = f^4 (1-f). 
\eeq
The Euler-Lagrange equation is given by 
\beq
\label{eq45}
{1 \over 2}f_{yy} + f - f^2 -{g^2 \over f^3} = 0. 
\eeq
For a given current $J$ (less than $J_c(T)$), there are two solutions to Eq. (\ref{eq44}).  The large f corresponds to the order parameter reduced in the presence of the current.  The smaller f solution corresponds to the maximum in the free energy.  Equation (\ref{eq45}) can be integrated once to lead to:

\beqn
\label{eq46}
{1 \over 4}{f_y}^2 + U(f) = K;~~~~~U(f) = {f^2 \over 2} - {f^3 \over 3} +{g^2 \over {2f^2}}. 
\eeqn

\begin{figure}[b]
\centering
\includegraphics[width=80mm]{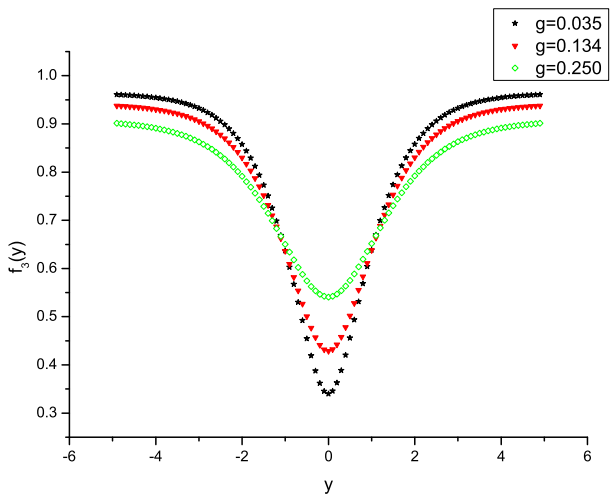}
%\hspace{0.0cm}\psfig{figure=f3_g.eps,height=8.5cm,width=10cm}
\caption{Order parameter profile $f_3 (y)$ for a third order phase transition in the presence of a current with $g = 0.035, 0.134$ and  $0.250$.}
\label{fig5}
\end{figure}

Here too, the solutions are those corresponding to the motion of a particle with position $f$, time $y$, moving in a potential $U(f)$.  The allowed values of $K$ are bounded by $K_1 < K < K_2$.  Here $K_1$ corresponds to the value of $U$ at the local minimum.  In the free energy this is the maximum corresponding to the smaller $f$ solution of Eq. (\ref{eq44}).  For $K \gtrsim K_1$, the solutions oscillate about the extremum $f_{max}$ with a wavelength determined by other parameters of the problem.  $K_2$ referes to the value of $U$ at the local maximum.  This corresponds to the free energy minimum and the order parameter as reduced by the current.  For $K \lesssim K_2$, the order parameter $f$ is constant everywhere except in a small region, where it drops down to a value $f_{min}$ such that $U(f_{min}) = K_2$.

Equation (\ref{eq46}) needs to be finally integrated.  The solution can be expressed as an integral. 

\beq
\label{eq47}
\int_{f_{min}}^f {{fdf} \over \sqrt{2f^5-3f^4+6Kf^2-3g^2}} = \sqrt{\frac{2}{3}} y. 
\eeq

\begin{figure}[t]
\centering
\includegraphics[width=80mm]{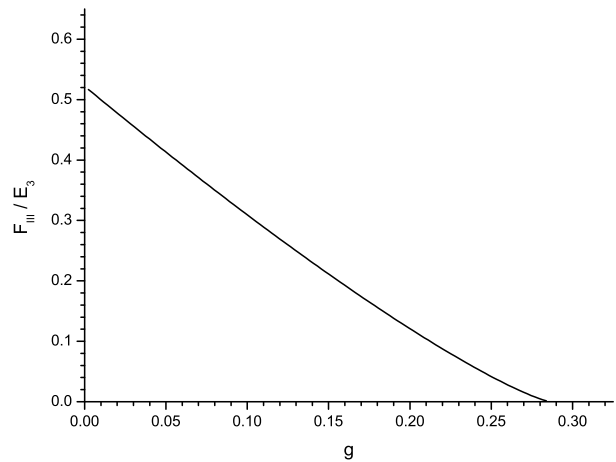}
%\hspace{0.0cm}\psfig{figure=e3_g.eps,height=6.5cm,width=6.5cm,angle=-90}
\caption{The energy of a phase slip center, numercailly interated from Eq.(\ref{eq43}) for a third order phase transition free energy.}
\label{fig6}
\end{figure}

The profile is shown in Fig.(\ref{fig5}).  Qualitatively the profile is similar\cite{la} to the one found for Eq. (\ref{eq12}).  The details however are all quite different.  Figure (\ref{fig6}) shows the energy of the phase slip centers as a function of the dimensionless supercurrent $g$ for the third order free energy.  When properly scaled, the results for a fourth order free energy look identical.  

A similar analysis can also be carried out for Eq. (\ref{eq14}), generalized to include the effect of a supercurrent.  In the dimensionless units the free energy becomes, 
\beq
\label{eq48}
\tilde{F}_{IV} = E_4 \int \left[-{1 \over 3}f^2 +{1 \over 4}f^{8/3} + {1 \over 9}{f_y}^2 - {g^2 \over {3f^2}}\right]. 
\eeq
Here the notation is identical to that following Eq. (\ref{eq26}).  The dimensionless current $g^2 = ({3J^2}/ {4 c_4})((4b)^6 / (3a)^7)$ is given by,
\beq
\label{eq49} 
g^2 = f^4(1-f^{2/3}). 
\eeq

One straightforward consequence of this equation (or Eq.(\ref{eq44})) is that the suppression of the order parameter due to a supercurrent is given here by $f^{2/3} \simeq 1-g^2$.  This and the result corresponding to Eq. (\ref{eq44}) $f \simeq 1-g^2$ look different only because $f$ depends differently on the order parameter.  Here the maximum in $g^2$ occurs (when the two solutions merge and there is no solution for larger g) at $f_o = (6/7)^{3/2} = 0.78$, corresponding to $g_c = 0.24$. 

We have not integrated the Euler-Lagrange equation for a phase slip center in this case.  In earlier calculations related to Eq. (\ref{eq14}), the final calculation seems to be possible only numerically.  Qualitatively the main features of the solutions for Eq. (\ref{eq13}) remain useful in offering insight into the solutions here.  We expect the profile of a phase slip center to follow the basic features of the solution described in Eq. (\ref{eq47}). The phase slip center will again appear as a spatially localized suppression of the order parameter, over a length scale which will vary with current.  Overall the energy of this structure will be $E_4 \xi$. 

\section{Summary and Conclusions}

In view of the success of a Landau free energy in describing issues connected with a second order phase transition, we have been stimulated to explore the properties of defects and textures in the presence of a higher order phase transition using generalized Landau free energy functionals.  We expect that, regardless of the so far unknown microscopic details, properties near a higher order transition can be described by a generalized free energy in essentially the same way that the usual Landau free energy captures the essential properties near a second order phase transition.

For a third order free energy, the transformation $f(y) = n^2(y)$ changes Eq.(\ref{eq13}) into Eq.(\ref{eq241}).  In the latter version, the order parameter must be positive also because otherwise the free energy is unbounded below.  That would eliminate the notion of $\exp (-f)$ as a configuration probability.  The constraint however leads to a very different mathematical structure for the defect profile.  In particular, the profile has a discontinuity in the slope $f_y$ at the origin. It is not altogether surprising that the positivity constraint leads to the discontinuity.  

For Eq.(\ref{eq14}), the transformed free energy is seen in Eq.(\ref{eq27}).  The term $f^{8/3}$ could be defined so that one is always taking the positive root.  There are, it seems no symmetry considerations, which will restrict $f$ and in this case one is left to speculate freely.  The microscopic theories will have more to say about whether the powers of the terms in the free energy are justifiable and the protocol for the proper approach to $f =0$.  The constraint $f>0$ has consequences for stability of the spatial textures which will be reported\cite{khare} separately.

There are defect and current related characteristic signatures of a higher order phase transition.  The current induced suppression of the order parameter is quadratic (in supercurrent) for small currents.  The critical current, at which the order parameter vanishes and superconductivity disappears, has a temperature dependence $J_c (T) \simeq a^{(2p-1)/2}$.  We have also calculated the energy and spatial profiles of structures such as an interface between degenerate superconducting states and a phase slip center which appears when the supercurrent approaches the critical current.

\section{Acknowledgements}
We are grateful to NSF and DOE for support. Discussions with have been valuable. 

%\newpage 

\end{document}